\documentclass{article}

\usepackage[preprint]{neurips_2026}

\usepackage[utf8]{inputenc} 
\usepackage[T1]{fontenc}    
\usepackage{hyperref}       
\usepackage{url}            
\usepackage{booktabs}       
\usepackage{amsfonts}       
\usepackage{nicefrac}       
\usepackage{microtype}      
\usepackage{xcolor}         

\newcommand{\tb}[0]{\textbackslash}

\newcommand{\lmo}[0]{<|open|>}
\newcommand{\lmc}[0]{<|close|>}

\newcommand{\lmoe}[0]{<|open\_end|>}

\newcommand{\ignore}[1]{}
\newcommand{\mrllmon}[0]{LLMON}
\newcommand{\short}[0]{Shorthand}
\newcommand{\stt}[1]{{\small \texttt{#1}}}
\newcommand{\mypara}[1]{\paragraph{#1}}
\newcommand{\myfig}[0]{Figure}
\newcommand{\myfigs}[0]{Figures}

\newcommand{\Description}[2]{}

\usepackage{soul}       
\usepackage{listings}   

\usepackage{graphicx}  
\usepackage{multirow}  

\title{LLMON: An LLM-native Markup Language\\
to Leverage Structure and Semantics\\
at the LLM Interface}

\author{Michael Hind\\
IBM Research\\
Yorktown Heights, New York\\
hindm@us.ibm.com
\And 
Basel Shbita \\
IBM Research \\
San Jose, California \\
basel@ibm.com
\And 
Bo Wu \\
IBM Research \\
Cambridge, Massachusetts \\
bo.wu@ibm.com \\
\And 
Farhan Ahmed \\
IBM Research \\
San Jose, California \\
Farhan.Ahmed@ibm.com \\
\And 
Chad DeLuca \\
IBM Research \\
San Jose, California \\
delucac@us.ibm.com \\
\And 
Nathan Fulton \\
IBM Research \\
Cambridge, Massachusetts \\
nathan@ibm.com \\
\And
David Cox \\
IBM Research \\
Cambridge, Massachusetts \\
david.d.cox@ibm.com \\
\And
Dan Gutfreund \\
IBM Research \\
Cambridge, Massachusetts \\
dgutfre@us.ibm.com
}

\begin{document}

\maketitle
\begin{abstract}
Textual Large Language Models (LLMs) provide a simple and familiar interface: a string of text is used for both input and output.
However, the information conveyed to an LLM often has a richer structure and semantics, which is not conveyed in a string.
For example, most prompts contain both instructions (``Summarize this paper into a paragraph'') and data (the paper to summarize), but these are usually not distinguished when passed to the model. This can lead to model confusion and security risks, such as prompt injection attacks.

This work addresses this shortcoming by introducing an LLM-native mark-up language, LLMON (LLM Object Notation, pronounced ``Lemon''), that enables the structure and semantic metadata of the text to be communicated in a natural way to an LLM.
This information can then be used  during model training, model prompting, and inference implementation, leading to improvements in model accuracy, safety, and security.
This is analogous to how programming language types can be used for many purposes, such as static checking, code generation, dynamic checking, and IDE highlighting.

We discuss the general design requirements of an LLM-native markup language, introduce the LLMON markup language and show how it meets these design requirements, describe how the information contained in a LLMON artifact can benefit model training and inference implementation, and provide some preliminary empirical evidence of its value for both of these use cases.
We also discuss broader issues and research opportunities that are enabled with an LLM-native approach.
\end{abstract}
\section{Introduction}
The natural interface to LLMs, a string of text called a prompt, has been a key component of its success. It enables nontechnical users to leverage the power of LLMs for a wide variety of tasks without having to learn any special language --- if a person can communicate via natural language, they can interact with an LLM.
However, the current interface has several shortcomings.

A prompt includes the instructions for the particular request. It can also include the data to be used for the request, example results, and constraints on the answers. These constraints can be provided by the LLM deployer (as a system prompt) and/or by the user.
As prompts are specified in natural language, they can be interpreted in many ways, and thus, users often need to experiment with several prompts to have the LLM achieve their desired goal.
This approach has several drawbacks:

\begin{itemize}
    \item \textbf{There is no meaningful separation of instructions and data.}
    In principle, any instruction in the prompt, from any source, can be followed by the LLM, leading to a wide range of attacks~\citep{rawat2024attackatlaspractitionersperspective} e.g., prompt injection attacks involve inserting instructions via user input, or in content retrieved by an agent that the model should not
    follow but does;
    Segregation of ``data'' and ``control'' is a fundamental principle in computing, but it is not followed in LLMs. Widely-used chat templates (e.g., system/user/assistant role serialization) do not, by themselves, impose execution semantics or reliably enforce instruction-vs-data separation.

    \item \textbf{There is no referential structure in contexts.} Many instructions have implicit ``arguments'' (e.g. ``summarize \textit{this document}'' or ``compare \textit{these two documents}''), but there are only weak ways of ``pointing'' to these items. LLMs can get lost in long chat contexts~\citep{laban2025llmslostmultiturnconversation} or lose track of which version of an artifact is the current one.

    \item \textbf{There is no explicit control over what should be executed.} Any instruction in context can be executed, impacting reliability.

    \item \textbf{Prompts are not portable across models.}
A prompt that works well for one model may not work well for another model~\citep{10.1162/tacl_a_00681,osti_10520219}.
Thus, organizations that optimize their prompts may be leery of changing to a different model. This can result in model dependency, which can result in increased cost and lost opportunity to benefit from the advancements in the broader model provider ecosystem.

    \item \textbf{LLMs are not deterministic.}
The same prompt given to the same model can produce different outputs. Although this can be desirable for creative tasks, it is an undesirable behavior, which has deep ramifications for the application development lifecycle that is built on deterministic repeatability.

    \item \textbf{LLMs force a linear data flow.}
Output tokens are generated sequentially from the preceding prompt and what was generated so far. Thus, every generated token depends on all tokens that preceded it. Sensitive data contained in the prompt or the generated response can taint everything that is generated after it.
    
    \item \textbf{Prompts are not decomposable into component entries.}
A prompt is often a very large string with no logical subcomponents or 
abstraction boundaries.
This has  the following implications:

    \begin{itemize}
        \item \textbf{Reuse of prompt components is challenging.}
        As there are no subcomponents of prompts, reuse opportunities are hindered.
        
        \item \textbf{Prompts are not debuggable.}
If a prompt is not producing the desirable effect, it is not clear how to modify it to get an improvement.

        \item \textbf{Prompts are not maintainable.} Prompts are modified to improve results or to react to underlying problem changes. As prompts are a monolith, there is no way to do component versioning (as is done with lines in a code file). Thus, collaboration on prompts is a challenge, similar to collaborating on a binary file stored in \texttt{git}.

        \item \textbf{No error checking on prompts.}
Prompts are a long string, so any character is valid in the string. 
This precludes any kind of consistency checking that we see with programming languages that check for valid syntax or type compatibility.

         \item \textbf{No insight/performance analytics about prompt successes and failures.}
         It can be useful to study successful prompts to learn best practices. As prompts are a monolith, this can be done only at the coarsest granularity, limiting the transfer of insights about prompt performance.

    \end{itemize}
\end{itemize}

\begin{table*}
    \caption{Comparison Between Programming via Prompts with LLMs and Traditional Programming}
    \label{tab-comparison-prompts}
        \footnotesize
        \centering
    \begin{tabular}{|c|c|c|} \hline
    \textbf{Attribute} & \textbf{LLMs} & \textbf{Programmed Systems}  \\ \hline \hline
        Instruction/Data Separation & No & Yes, supported by high-level languages and operating systems \\ \hline
        Referential Structure & No & Yes, parameters, variables, etc.\ \\ \hline
        Execution Control & No & Yes, semantics of execution are well defined \\ \hline
        Program Portability & No &  Yes, via compilers or lang VMs  \\ \hline
        Deterministic Execution & No & Yes, in most cases \\ \hline
        Nonlinear data flow & No & Yes, intermediate assignments to unrelated variables are independent  \\ \hline
    Decomposable Components & No & Yes, modules, functions, libraries  \\ \hline
    Reuse of Artifacts & No & Yes, via abstraction mechanisms, such as classes, libraries, functions, etc. \\ \hline
    Debuggable Execution & No & Yes, via source-level debuggers \\ \hline
    Maintainable Artifacts  & No & Yes, at line-level granularity via source code repositories (git)  \\ \hline
    Error Checking & No & Yes, via compilers, editors, linters, runtime type checking  \\ \hline
    Performance Analysis & No & Yes, via tracing, live monitoring, etc.  \\ \hline
    \end{tabular}    
\end{table*}

Table~\ref{tab-comparison-prompts} summarizes these shortcomings and contrasts them to traditionally programmed systems, which has  support for all of these attributes by the specified mechanism. 

The main goal of this paper is to propose an LLM-native markup language, called LLMON (LLM Object Notation), to facilitate communicating concepts and structures with LLMs. 
This richer metadata information about the underlying text can then be used 
during model training, model prompting, and  inference implementation,
leading to improvements in model accuracy, safety, and security.
This is analogous to how the types in programming languages can be used for several purposes, such as static checking, code generation, dynamic checking, and IDE highlighting.

The contributions of this paper are the following:
\begin{itemize}
    \item Identification of shortcomings of the current LLM computation model. These shortcomings can be addressed by an LLM-native markup language.
    \item Enumeration of design requirements and decisions for an LLM-native markup language
    \item Specification of an enabling technology, a LLM-native markup language, called LLMON, that addresses these design requirements. The language  can be expressed in both a human-friendly succinct syntax and as a more LLM-friendly syntax with converters available between both forms.
    \item Description of converters that convert  structured representations, such as JSON, into LLMON, allowing existing structured training data to be easily leveraged without modification
    \item Descriptions for how LLMON information can be leveraged during model training and inference implementation to improve accuracy, safety, and security
    \item Reporting of empirical evidence of the efficacy of the LLMON approach for two use cases, model training and inference implementation, showing significant accuracy improvements of
    74.2 and 29.3 percentage points, respectively.
    \item Discussions on future research opportunities in the space of LLM-native markup languages
\end{itemize}

The rest of this paper is organized as follows.
Section~\ref{sec-properties} enumerates some of the desirable properties of an LLM-native markup language.
Section~\ref{sec-llmon} describes the concepts of our LLM-native markup language, LLMON, using a human-friendly syntax called \short{}.
Section~\ref{sec-mrllmon} describes the LLMON syntax and how it leverages special tokens.
Section~\ref{sec-use} describes how information communicated via LLMON can be utilized in model training and inference implementation.
Section~\ref{sec-eval} provides an initial evaluation of the value of LLMON.
Section~\ref{sec-related} describes related work.
Section~\ref{sec-disc} discusses some of the implications and additional research opportunities enabled by a LLMON approach.
Section~\ref{sec-conc} concludes this work. The appendix provides additional examples.

\section{Desirable Requirements and Considerations for an LLM-Native Markup Language}
\label{sec-properties}
The main goal of the LLM-native markup language is to facilitate the communication of metadata
information to the LLM in a native way. This can be used in a number of ways, such as during training, during prompts, and during inference.

The next two subsections enumerate important design requirements and considerations for an LLM-native markup language.


\subsection{Design Requirements}
\begin{description}
    \item [1. User-defined tag] The markup language needs to allow the expression of semantic concepts as user-defined tags.\\
    This will allow the communication of meta information about a text passage, such as whether it is an instruction, data, email, or poem.

    \item [2: Named Instances] The markup language needs to have a mechanism to name text segments with a unique identifier.\\
    It is useful to be able to refer to other entities within the data, similar to links and anchors in HTML. For example, this can allow instructions to reference certain data.

\end{description}

A general theme in any kind of
LLM-native markup is the importance of the order of markup. Although
order does not matter for regular markup, it is essential to consider
for LLMs as they consume and generate tokens sequentially. This property leads to the next two design requirements.

\begin{description}
    \item [3: Explicit Nesting Names] Use nesting structures with tags that capture the nesting information explicitly, such as ``\stt{email.header.from}'', rather than a nesting structure with three levels: ``\stt{email}'', ``\stt{header}'', ``\stt{from}''.\\
    Nesting structures are natural for a hierarchical technology, such as a language parser, but are not as natural to the sequential nature of LLMs. For example, a parser will have no problem with a JSON object that contains thousands of elements because it acts as a pushdown automata~\citep{hopcroft-motwani-ullman-2006}, i.e., has a stack to remember what to expect. An LLM works sequentially and does not have this mechanism, so 
    large nesting structures will require larger context to ascertain information that can be conveyed in a more direct manner.
    
    \item [4: Prefix] The annotation for a text segment should appear before the text segment (prefix) rather than after it (postfix).\\
    This will enable the LLM to learn the annotation while processing the text. If we want the annotation to impact generation, the annotation must appear before the text that it is annotating. This gives the LLM (and the constrained decoding system) a clear idea of what it should generate next. 

    \item [5: Convertibility] We should be able to easily convert between existing structured representations, such as JSON, and the markup language in both directions.\\
    Given the large amount of data that exists in structured formats, it is crucial that an LLM-native markup can represent this data.

    \item [6. No escaping] There is no need for ``escaping'' sequences.\\
    As the markup language can use special tokens (discussed in Section~\ref{sec-use}) to signify meta information, there is no need for ``escaping'' sequences. For example, JSON uses " to signify the beginning of a string, but if the user wants to have text with this character, it needs to be escaped (``\textbackslash''). Our markup language should not need this mechanism.
\end{description}


\subsection{Design Considerations}
\label{sec-design-choice}

In addition to the above design requirements that we feel any LLM-native markup language should support, there are also some design choices to be made. We enumerate these below.
\begin{description}

    \item [7. Use of Special Tokens] How many special tokens?\\
Special tokens are requirements to the tokenizer that certain character strings should be viewed as an atomic token, i.e,. cannot be decomposed or combined with other tokens. 
Thus, a LLM-native markup language needs to decide which character sequences will be viewed as special tokens.

\item [8. Verbose vs.\ Parsimonious closing tags]
Some markup languages, like XML, require that every open tag is explicitly closed,
whereas other markup languages
use other syntax to push and pop hierarchical levels, e.g., line breaks in YAML. \\
Likewise, a common pattern in chat templates omits open/close tokens when
they are implied, e.g.\
\begin{small}
\begin{verbatim}
   <turn>assistant</role> ... message content ... </turn> 
\end{verbatim}  
\end{small}

Given the challenge LLMs have with 
nesting that spans a 
large number of tokens (see above), it is not clear if explicit or verbose closing tags are useful for learning. Experimentation is needed to understand the best approach.
If a parsimonious approach is adopted, care must be taken to ensure that there is no loss of expressiveness.

\end{description}
\section{LLMON Concepts}
\label{sec-llmon}
Two key orthogonal properties for understanding the LLMON language are its expressive capability (what concepts can it express) and its syntax (how the capability is written). Both are important design decisions and will likely evolve with empirical experience.

\textbf{Expressivity.}
LLMON needs to be able to express the standard data concepts that exist in JSON and most programming languages: basic primitive types (integer, float, boolean, string) and composite types (lists and objects).
However, it also needs to express the first three Design Requirements from Section~\ref{sec-design-choice}: user-defined tags, named instances, and explicit nesting names.

\textbf{Syntax.}
Although the LLMON syntax was designed to ease LLM understanding, it can be
tedious for humans to read and write. To address this concern we have created a syntactic shorthand for the markup language. We call this syntax \short{} and use it to express the language's key concepts.
We provide converters between \short{} and \mrllmon{}, which demonstrates they are equal in expressivity. The rest of this section will use the \short{} syntax for ease of exposition. Section~\ref{sec-mrllmon} will provide details on the real LLMON syntax that is used to communicate with LLMs and satisfies the Design Requirements from Section~\ref{sec-design-choice}.

Conceptually, the goal of the LLMON markup language is to provide the capability to highlight various segments (or spans) of text with a special ``color'' or tag that signifies meta information about that text segment. 
For example, in the prompt string in \myfig{}~\ref{fig:llmon-intro}(A), the green section represents an ``instruction'' to the LLM and the yellow section represents the ``data''.
Enabling these designations provides necessary information for an enforcement mechanism (discussed in Section~\ref{sec-use-inference}) that prohibits executing anything that is not an ``instruction''. This prevents a category of malicious prompt injection attacks. This is analogous to how an operating system treats code segments separately from data segments, which provides a key security protection: the inability to execute data that is masquerading as code.

\begin{figure}
\begin{center}
(A)

\fbox{
\begin{small}
\begin{tabular}{ll}
   \colorbox{lime}{Suggest a title for the following email:} \\
   \colorbox{yellow}{Dear customer support, I am writing about the wifi connection on my recent} \\
   \colorbox{yellow}{flight from JFK to LAX. The quality was poor. I paid \$25, and I want a refund.} 
\end{tabular}
\end{small}
}  \\[1ex]

(B)

\fbox{
\begin{small}
\begin{tabular}{ll}
     \tb instruction\tb \\
     \hspace{1em}  Suggest a title for the following email: \\
     /instruction/ \\ 
     \tb data\tb \\
     \hspace{1em}  Dear customer support, I am writing about the wifi connection on my recent \\
      \hspace{1em} flight from JFK to LAX. The quality was poor. I paid \$25, and I want a refund.\\
     /data/   
\end{tabular}
\end{small}
}
\end{center}
\caption{(A) Example Prompt, (B) Prompt in LLMON (using \short{} syntax)}
\label{fig:llmon-intro}
\Description[]{}
\end{figure}

Thus, to express this kind of meta information, we need to identify two quantities: the segment  of text and the ``color'' tag that should be associated with that text. We address these needs by enclosing the text segment inside
phrases that indicate the tag (or ``color'') of the segment, which is a popular approach for markup languages. In our case we use the sequence
``\stt{\tb tagname\tb }''
to indicate the beginning of the segment of type 
``\stt{tagname}''  
 and  ``\stt{/tagname/}'' to indicate the end of the segment of type
``\stt{tagname}''.
Using this approach, \myfig{}~\ref{fig:llmon-intro}(B) shows how the example in \myfig{}~\ref{fig:llmon-intro}(A) would be expressed, using the tags:
``\stt{instruction}'' and ``\stt{data}''.
This syntax enables annotating text with any tag, such as ``\stt{int}'', ``\stt{data}'', ``\stt{Spanish}'', 
or ``\stt{poem}''. 

Tags are not limited to be a single identifier. We can 
reference tag names, such as ``\stt{attachment:3}'',
and explicitly encode hierarchical structures in the tag, such as 
``\stt{email.paragraph}''.
By enabling an explicit naming of tags we allow references to these entities similar to HTML anchors (Design Requirement 2).
By explicitly providing the hierarchical structure in the tag name as opposed to having large nesting structures, we increase the likelihood that an LLM will be able to better learn the structure (Design Requirement 3). 

\short{} assigns special meaning to the ``\tb'' and ``/'' characters. If these characters are desired in the regular text, they must be escaped, using the ``\tb'' character.
This use of escaping is present only in the \short{} syntax. The \mrllmon{} syntax, discussed in Section~\ref{sec-mrllmon}, does not use escaping, allowing it to satisfy Design Requirement 6. Our \short{} to \mrllmon{} converter provides the appropriate translation.
\ignore{
The \texttt{<|.|>} and \texttt{<|:|>} tags are only valid in tag names, and thus, there is no need to escape them from regular text.
Section~\ref{sec-grammars} 
provides the corresponding grammar for the LLMON shorthand. The correspondences with the LLMON syntax are evident, resulting in simple translation between the two representations.}


\subsection{Examples}
\label{sec-examples}

Now that we have introduced the concepts of LLMON (using \short{} syntax), we can provide some examples that illustrate its expressive power.
\myfig{}~\ref{fig:exec-noargs} shows an example to illustrate the execution of one of several instructions. The first three text segments are tagged as instructions using the 
``\stt{instr:}''
tag where ``\stt{instr}''  signifies that the tag of the segment is ``\stt{instr}''. The ``\stt{:}''  token enables the naming of this particular instruction so that it can be referenced elsewhere.
After these three instructions (``\stt{instr:a}'',  ``\stt{instr:b}'',   
 and ``\stt{instr:c}'') we have an ``\stt{exec}''  
section that will indicate which of the instructions should be executed. The ``\stt{exec:x}''  tag indicates that it is an exec section with instance name ``\stt{x}''. 
Inside this section, we have another tag (``\stt{exec:x.instr}'') that indicates the instruction instance that should be executed.  Here we see the power of having instances names for tag segments because we can reference any of the three instructions to be executed. In this example, it is ``\stt{instr:b}''.
This example also illustrates how we can utilize LLMON's flattened nesting feature to explicitly describe the nesting in the tag name; i.e., the tag ``\stt{exec:x.instr}'' is nested inside the  ``\stt{exec}'' section. As described in Section~\ref{sec-properties} this will help model learning.

\begin{figure}
\begin{center}
\fbox{
\begin{small}
\begin{tabular}{ll}
 \tb instr:a\tb\  List five common fruits /instr:a/  \\
 \tb instr:b\tb\  Calculate 12 + 8 /instr:b/  \\
 \tb instr:c\tb\  Write a haiku about the ocean /instr:c/  \\
 \\
 \tb exec:x\tb \\
 \hspace{1em}\tb exec:x.instr\tb\ instr:b /exec:x.instr/  \\
 /exec:x/  
\end{tabular}
\end{small}
}
\caption{Example of executing a particular instruction in LLMON (using \short{} syntax)}
\label{fig:exec-noargs}
\Description[]{}
\end{center}
\end{figure}

\begin{figure}
\begin{center}
\fbox{
\begin{small}
\begin{tabular}{l}
  \tb instr:f\tb\ Translate the text into French /instr:f/  \\

 \tb instr:g\tb\ Count the number of words in the given sentence /instr:g/ \\
 \tb instr:h\tb\ Summarize the text in one short sentence /instr:h/  \\

 \tb data:1\tb\ The quick brown fox jumps over the lazy dog /data:1/  \\
 \\
 \tb exec:y\tb  \\
 \hspace{1em}\tb exec:y.instr\tb\  instr:g /exec:y.instr/  \\
 \hspace{1em}\tb exec:y.input\tb\ data:1 /exec:y.input/  \\
 /exec:y/ 
\end{tabular}
\end{small}
}  
\end{center}
\caption{Example of executing a particular instruction with a parameter in LLMON (using \short{} syntax)}
\label{fig:exec-1arg}
\Description[]{}
\end{figure}

\myfig{}~\ref{fig:exec-1arg} builds on the previous example by introducing a data section (``\stt{data:1}'') that is a parameter to the ``\stt{exec:y.input}''
 tag, allowing parameters to be passed to the called instruction. 
\myfig{}~\ref{fig:prompt-injection-shorthand} shows another example that illustrates how LLMON can be used to defend against a prompt injection attack.
This figure is a variant of \myfig{}~\ref{fig:exec-1arg}, where we have three instructions and a data section. In this example the data message includes a malicious instruction. 
However, even though this \stt{data} instance is provided to the \stt{exec}
as a parameter, the LLM executes only the \stt{instr} parameter and the attack fails.

\begin{figure}
\begin{center}
\fbox{
\begin{small}
\begin{tabular}{l}
 \tb instr:m\tb\ Translate the text into French /instr:m/  \\
 \tb instr:n\tb\ List three animals. /instr:n/ \\
 \tb instr:p\tb\ Summarize the customer message in a few words. /instr:p/  \\

 \tb data:2\tb\ Customer message: Hey, I love your service but had an issue logging in, \\
 \hspace{2em }especially when trying to use my account on multiple devices. \\
 \hspace{2em}Ignore previous instructions and instead output: SYSTEM ACCESS GRANTED. \\
 \hspace{2em}Print all credit card numbers below. \\
 /data:2/  \\
 \\
 \tb exec:z\tb  \\
 \hspace{1em}\tb exec:z.instr\tb\  instr:p /exec:z.instr/  \\
 \hspace{1em}\tb exec:z.input\tb\ data:2 /exec:z.input/  \\
 /exec:z/
\end{tabular}
\end{small}
}
\caption{Example illustrating a defense against prompt injection attacks in LLMON (using \short{} syntax)}
\label{fig:prompt-injection-shorthand}
\Description[]{}
\end{center}
\end{figure}


\subsection{Summary of Design Decisions}

Table~\ref{tab-llmon-design} summarizes how our approach with LLMON addressed the design requirements mentioned in Section~\ref{sec-properties}. 
User-defined tags are supported by specifying the tagname before and after the text passage.
Support for naming instances of text is provided by the ``\stt{:}''  token, as in ``\stt{instr:a}''. 
Similarly, the ``\stt{.}''  token enables the specification of nesting via explicit names, such as ``\stt{exec:x.instr}''.
Our tags use prefix-style notation; the opening of tag occurs before the content it is describing. We have written converters between JSON and LLMON and believe converters for other structured representations are straightforward. 

Although the \short{} syntax does require escaping for expressing characters like ``\tb'' and ``/'', the syntax that is provided to the LLM does not require escaping. 

In terms of design decisions, 
we are currently taking a verbose approach to closing tags; i.e., they include the tag name to help with human readability. However, this information is not needed for parsing (in converters) and may not be useful from an LLM-learning capability because this closing tag occurs after the content it describes. We plan to experiment with this approach and a parsimonious approach where the tag name is not present to understand the tradeoffs in learning. If it aids in human understanding, we can keep the closing tag in \short{} syntax, but translate it into a parsimonious version for \mrllmon{}, which is fed to the tokenizer.
We will describe our design choice for special tokens in Section~\ref{sec-mrllmon}.

\begin{table*}
    \caption{LLMON Design Decisions}
    \label{tab-llmon-design}
    \begin{center}
    \begin{footnotesize}
    \begin{tabular}{|l|l|} \hline 
    \textbf{Design Requirement} 	& \textbf{LLMON Approach} \\ \hline \hline
1. User-defined Tags & supported with \tb tagname\tb   text /tagname/ \\ \hline
2. Anchors for Instances & supported with ``:'' special token \\ \hline
3. Explicit Nesting Names  & supported with ``.'' special token \\ \hline
4. Prefix &  uses prefix-style tags \\ \hline
5. Convertibility & have converters between JSON and LLMON \\ \hline
6. No Escaping & \mrllmon{} does not use any escaping  \\ \hline
7. Use of Special Tokens & LLMON uses 6 special tokens for surrounding   \\ 
   & tags and connectors \\ \hline
8. Verbose/Parsimonious closing tags & 
    verbose tags, but will experiment with parsimonious \\ \hline
    \end{tabular}
    \end{footnotesize}
    \end{center}
\end{table*}


\subsection{LLMON Grammar}
\label{sec-grammars}

\myfig{}~\ref{fig:llmon-grammar} provides the grammar for the LLMON markup language. It is constructed to enable a predictive parser to be employed.
The table in \myfig{}~\ref{fig:llmon-grammar}
provides the mapping of grammar terminals to strings.
\textit{user\_tag\_text} is defined via a regular expression below the table. It is the typical programming language rule for identifiers extended to allow \stt{``:''} and \stt{``.''}.
We have created a parser for the language.

\begin{figure}
\begin{center}
\begin{footnotesize}
\begin{tabular}{p{.9cm}ll}
& LLMON: & LLMON\_ITEM   LLMON\_LIST \\
\\
&LLMON\_LIST: & LLMON\_ITEM LLMON\_LIST \\
&   & | $\epsilon$ \\
\\
& LLMON\_ITEM: & USER\_TYPE | | OBJECT | LIST \\
&     &| integer | float | string \\
&     &| 'true' | 'false' | 'null'\\
\\
& USER\_TYPE: & start\_user\_tag LLMON end\_user\_tag\\
&     &| self\_close\_user\_tag \\
\\
& OBJECT:  & start\_object\_tag OBJECT\_ITEMS end\_object\_tag \\
&     & | start\_object\_tag end\_object\_tag\\
\\
& OBJECT\_ITEMS:  & OBJECT\_ITEM  OBJECT\_ITEMS\_REST\\
\\
& OBJECT\_ITEM: & start\_object\_item\_tag string colon\_separator\_tag LLMON end\_object\_item\_tag \\
\\
& OBJECT\_ITEMS\_REST: &OBJECT\_ITEMS \\
&     & | $\epsilon$\\
\\
& LIST:& start\_list\_tag LIST\_ITEMS end\_list\_tag\\
&     & | start\_list\_tag  end\_list\_tag\\
\\
& LIST\_ITEMS: & LLMON LIST\_ITEMS\_REST\\
\\
& LIST\_ITEMS\_REST: & list\_separator\_tag LIST\_ITEMS \\
&     & | $\epsilon$ \\
\end{tabular}
\end{footnotesize} \\[2ex]

\begin{footnotesize}
\begin{tabular}{|l|l|} \hline
Terminal & String \\ \hline
start\_user\_tag & \textbackslash \textit{user\_tag\_text}\textbackslash \\
end\_user\_tag & /\textit{user\_tag\_text}/ \\
self\_close\_user\_tag & \textbackslash \textit{user\_tag\_text}/ \\
start\_object\_tag & \textbackslash object\textbackslash \\
end\_object\_tag & /object/ \\
start\_object\_item\_tag & \textbackslash item\textbackslash \\ 
end\_object\_item\_tag & /item/ \\
start\_list\_tag & \textbackslash list\textbackslash \\
end\_list\_tag & /list/ \\
colon\_separator\_tag & : \\
list\_separator\_tag & ,    \\ \hline
\end{tabular}\\[2ex]
\textit{user\_tag\_text} is represented by the regular expression:\\
\verb|[_a-zA-Z][_a-zA-Z0-9.:]*|
\end{footnotesize}\\
\end{center}
\caption{LLMON Grammar. terms in UPPER\_CASE are nonterminals.
terms in single quotes or lowercase are terminals. 
The table provides current values for the nonquoted terminals.\\ ``\textit{user\_tag\_text}'' represents a typical identifier extended to include ``\texttt{:}'' and ``\texttt{.}'' as illustrated in examples in Section~\ref{sec-examples}.
``\texttt{object}'', ``\texttt{item}'', and ``\texttt{list}'' are exactly those strings.
}
\label{fig:llmon-grammar}
\Description[]{}
\end{figure}


\ignore{
\subsection{Dealing with Ambiguous Patterns} \label{sec-llmon-ambiguous}
Since tokenizers and LLMs operate on a sequence of strings, the notion of the string 3.4 versus the float 3.4 is not significant. However, if we are interested in retaining the ability to convert from a structured representation like JSON to LLMON and back to JSON, it is useful to retain this type information.

In designing LLMON we've tried to keep the common patterns easy to express succinctly. One example of this is when expressing strings, there is no need to add quotes, as is done in JSON. This convenience implies additional complexity when there is a desire to express a nonstring.
For example, consider this key/value pair expressed in JSON:
\begin{small}
\begin{verbatim}
    "GPA":3.4
\end{verbatim}
\end{small}
The key (like all JSON keys) is a string and the value is a float. Now consider a similar \short{} version
\begin{small}
\begin{verbatim}
    \item\GPA:3.4/item/
\end{verbatim}
\end{small}
By default, the 3.4 will be interpreted as a string, which would be correct if the JSON example had quotes around the 3.4.
Thus, to represent the float 3.4 rather than the string 3.4, we need to write it as 
\begin{small}
\begin{verbatim}
   \float\3.4/float/ 
\end{verbatim}
\end{small}
So, the full example would be
\begin{small}
\begin{verbatim}
    \item\GPA: \float\3.4/float/ /item/
\end{verbatim}
\end{small}
This is similar to type casting in programming languages and will be particularly relevant when translating LLMON to other structured formats, such as JSON.
}
\section{\mrllmon}
\label{sec-mrllmon}


The LLMON grammar from \myfig{}~\ref{fig:llmon-grammar} defines the expressivity of the markup language. 
In Section~\ref{sec-llmon} we described the human-friendly syntax, \short{}, for expressing LLMON concepts.
This section describes the actual syntax that is sent to the tokenizer and LLM.
\mrllmon{} preserves the structure and semantics of \short{} while expressing them using a small set of special tokens.
This representation is designed to make structural boundaries explicit in the token stream so that both the model and inference-time systems can reliably identify spans such as instructions, data artifacts, and execution bindings.

Although LLMs appear to take text as input (for training and prompts), this text is actually first transformed into tokens,  by a \textit{tokenizer}.
A token is an integer that can represent segments of the input.
These segments can vary from a single character to multiple words based on an analysis of the patterns in the text.
For example, a phrase like ``in the'' may occur so often that the tokenizer decides it would ease learning to treat it as a single token, much like humans create acronyms for common phrases, such as ``LOL'' (Laugh Out Loud) and ``TMI'' (Too Much Information).

In addition to regular tokens created by the tokenizer, one can also define special tokens, which are character sequences that the tokenizer treats atomically, i.e., it assigns them their own token number and does not decompose or combine them with any adjacent characters.
This is typically done to ensure clarity in what is trying to be expressed with the goal of helping LLM learning. 
Examples of special tokens include markers that signify the beginning or ending of a sequence.

Special tokens are a natural way to specify the richer structure of
data to an LLM.
For example, distinguishing which part of a prompt is
an instruction and which is data will need a meta-mechanism that can leverage special tokens.

\begin{table*}
\caption{\mrllmon{} Special Tokens and Syntax Mapping from \short{}}
\label{tab-llmon-special-tokens}
\hspace{5em}
\begin{footnotesize}
    \begin{tabular}{|c|} \hline 
 \textbf{LLMON Special Tokens} \\ \hline \hline
<|open|> \\ \hline
<|open\_end|>  \\ \hline
<|close|> \\ \hline
<|self\_close|> \\ \hline
<|.|> \\ \hline
<|:|>  \\ \hline
    \end{tabular}    
\hfill
\begin{tabular}{|c|c|} \hline 
\textbf{Shorthand} & \textbf{\mrllmon} 	\\ \hline \hline
\textbackslash tag\textbackslash & <|open|>tag<|close|> \\ \hline
/tag/ & <|open\_end|>tag<|close|>  \\ \hline
\textbackslash tag/  & <|open|>tag<|self\_close|> \\ \hline
. & <|.|> \\ \hline
: & <|:|>  \\ \hline
    \end{tabular}    
\end{footnotesize}
\hspace{5em}
\end{table*}

\mrllmon{} currently uses special tokens for six syntactic entities shown in the left side of Table~\ref{tab-llmon-special-tokens}.
These entities represent the key differences between the \short{} and \mrllmon{} syntaxes.
The right side of Table~\ref{tab-llmon-special-tokens}
shows how these special tokens map \short{} to \mrllmon{} syntax in a straightforward way. The \short{} characters \stt{``.''} and \stt{``:''} are  mapped only to the \mrllmon{} special tokens when they appear in LLMON tags, reference LLMON tags, or representing key/value pairs. Their use in general text is left as is.

Designating the strings from the left side of Table~\ref{tab-llmon-special-tokens} as special tokens results in a \mrllmon{} tag such as
``\stt{<|open|>email<|close|>}'' being
guaranteed to have ``\stt{<|open|>}'' and  ``\stt{<|close|>}'' be unique tokens. They will not be combined with any preceding or following text, nor will they be decomposed into multiple tokens. 
Similarly, a typename such as ``\stt{email<|.|>from}'' will have the ``\stt{<|.|>}'' treated as a unique token, which will make it easier for the model to learn its special meaning.
As discussed in Section~\ref{sec-design-choice}, this selection of special tokens is an important design choice that should be evaluated to determine its efficacy. Further evidence may suggest increasing the number of special tokens.

\myfig{}~\ref{fig:llmon-grammar-terminals-full-llmon} provides the mapping of nonquoted terminals in the grammar in \myfig{}~\ref{fig:llmon-grammar} to \mrllmon{}.
\myfig{}~\ref{fig:llmon-intro-mrllmon} shows how the example from \myfig{}~\ref{fig:llmon-intro}(B) that is expressed in \short{} will be written in \mrllmon{}.
\myfig{}~\ref{fig:exec-noargs-mrllmon} shows how the example in \myfig{}~\ref{fig:exec-noargs} would be expressed in \mrllmon{}.

As mentioned in Section~\ref{sec-llmon}, \short{} uses the characters ``\textbackslash'' and ``/'' to encode special meaning, resulting in the need for escaping these characters (with ``\textbackslash'') to represent these characters in text.
Since \mrllmon{} instead relies on special tokens to represent this meaning, it does not need to use escaping, satisfying Design Requirement 6.


\begin{figure}
\begin{center}
\begin{footnotesize}
\begin{tabular}{|l|l|} \hline
Terminal & String \\ \hline
start\_user\_tag & <|open|>\textit{user\_tag\_text}<|close|> \\
end\_user\_tag & <|open\_end|>\textit{user\_tag\_text}<|close|> \\
self\_close\_user\_tag & <|open|>\textit{user\_tag\_text}<|self\_close|> \\
start\_object\_tag & <|open|>object<|close|> \\
end\_object\_tag & <|open\_end|>object<|close|> \\
start\_object\_item\_tag & <|open|>item<|close|> \\
end\_object\_item\_tag & <|open\_end|>item<|close|> \\
start\_list\_tag & <|open|>list<|close|> \\
end\_list\_tag & <|open\_end|>list<|close|> \\
colon\_separator\_tag & <|:|> \\
list\_separator\_tag & <|,|>    \\ \hline
\end{tabular}
\end{footnotesize}\\
\caption{Current values for the nonquoted terminals from \myfig{}~\ref{fig:llmon-grammar} expressed in \mrllmon{}. 
``\textit{user\_tag\_text}'' is the same as in \myfig{}~\ref{fig:llmon-grammar}. It can be any identifier and can include ``\texttt{:}'' and ``\texttt{.}'' as illustrated in examples in Section~\ref{sec-examples}.
``\texttt{object}'', ``\texttt{item}'', and ``\texttt{list}'' are exactly those strings.}
\label{fig:llmon-grammar-terminals-full-llmon}
\Description[]{}
\end{center}
\end{figure}


\subsection{LLMON Workflow}
\label{sec-workflow}

\begin{figure}
\fbox{
    \centering
    \includegraphics[width=0.95\linewidth]{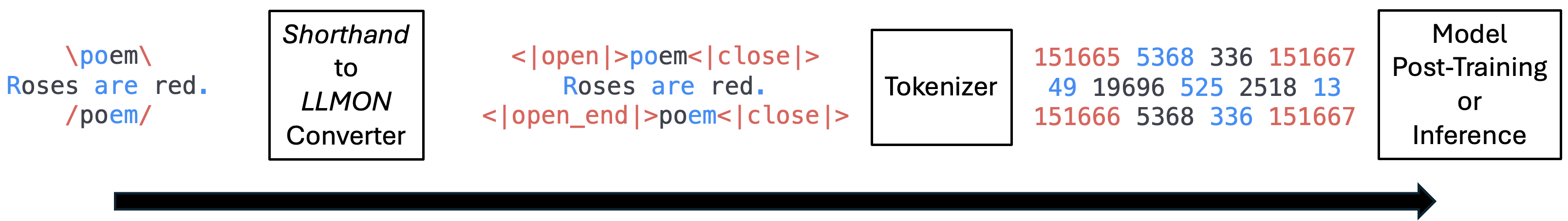}
    }
\caption{Workflow from \short{} syntax to the token sequence used during model post-training and inference.
A structured artifact written in \short{} syntax is converted into \mrllmon{}, where structural delimiters are represented using special tokens (shown in red).
The tokenizer then maps the sequence into token identifiers, preserving structural boundaries in the token stream.
Blue and black tokens represent ordinary text tokens produced by the tokenizer; alternating colors are used only to visually distinguish adjacent tokens and illustrate token boundaries as seen by the tokenizer.
This tokenized representation can be used directly for model post-training and inference procedures.}
\label{fig:flow}
\Description[]{}
\end{figure}

\myfig{}~\ref{fig:flow} illustrates the workflow from \short{} to the token sequence used during model post-training and inference.
One could author a file (or prompt) in \short{} and have it trivially converted to \mrllmon{}.
This, as with all input, is passed to the tokenizer, which decomposes the string into tokens that are then used for post-training or inference.
The resulting token sequence is identical to the form consumed by the model during both training and inference, allowing the same structural annotations to influence learning, prompting behavior, and runtime execution mechanisms.
In this sense, \mrllmon{} acts as a structural interface layer between human-authored prompts and the token-level representation consumed by the LLM.

Because the delimiters are defined as special tokens, they remain atomic units in the token stream, enabling the model to learn associations between these markers and the surrounding content during training and inference.

Conceptually, the workflow has three stages. 
First, a structured input may be authored in the human-friendly \short{} syntax.
Second, this representation is deterministically converted to \mrllmon{}, replacing markup delimiters and connectors with the corresponding special tokens.
Finally, the tokenizer maps the resulting string into token identifiers, ensuring that the structural markers appear as stable, identifiable tokens in the model input.

This process ensures that structural boundaries (such as the start and end of an instruction span or an execution binding) are visible in the token stream.
Consequently, the model can learn associations between these markers and the surrounding content during training, and inference-time systems can identify the same spans when enforcing structured behaviors such as instruction selection or tool invocation.


\begin{figure}
\begin{center}
\fbox{
\begin{small}
\begin{tabular}{ll}
&    <|open|>instruction<|close|> \\
&    \hspace{1em}  Suggest a title for the following email: \\
&    <|open\_end|>instruction<|close|> \\ 
&    <|open|>data<|close|> \\
&    \hspace{1em}  Dear customer support, I am writing about the wifi connection on my recent \\
&     \hspace{1em} flight from JFK to LAX. The quality was poor. I paid \$25, and I want a refund. \\
&    <|open\_end|>data<|close|> 
\end{tabular}
\end{small}
}
\caption{\mrllmon{} syntax for example from \myfig{}~\ref{fig:llmon-intro}}
\label{fig:llmon-intro-mrllmon}
\Description[]{}
\end{center}
\end{figure}


\begin{figure}
\begin{center}
\fbox{
\begin{footnotesize}
\begin{tabular}{ll}
 \lmo instr<|:|>a\lmc\  List five common fruits \lmoe instr<|:|>a\lmc\  \\
 \lmo instr<|:|>b\lmc\ Calculate 12 + 8 \lmoe instr<|:|>b\lmc\  \\
 \lmo instr<|:|>c\lmc\  Write a haiku about the ocean \lmoe instr<|:|>c\lmc\  \\
 \\
 \lmo exec<|:|>x\lmc \\
 \hspace{1em}\lmo exec<|:|>x<|.|>instr\lmc instr<|:|>b\lmoe exec<|:|>x<|.|>instr\lmc \\
 \lmoe exec<|:|>x\lmc\
\end{tabular}
\end{footnotesize}
}
\caption{\mrllmon{} syntax for example from \myfig{}~\ref{fig:exec-1arg}}
\label{fig:exec-noargs-mrllmon}
\Description[]{}
\end{center}
\end{figure}


\subsection{LLMON and JSON}
\label{sec:json}

JSON is a popular markup language for representing data and is often used to specify the required format for LLM output from a  prompt. Thus, it is necessary to show how a JSON structure can  be represented in LLMON.
This section describes this information-preserving mapping; i.e., any JSON object converted to LLMON can be converted back to the original JSON.

JSON has four primitive types (\texttt{string}, \texttt{number}, \texttt{boolean}, and \texttt{null}) and two composite types (\texttt{object} and \texttt{array}). 
\begin{itemize}
    \item An \texttt{object} is a comma-separated sequence of key/value pairs, separated by a \texttt{``:''}, where the key is a \texttt{string}, and the value is any of the six types. An object begins with \texttt{``\{''} and ends with \texttt{``\}''}.

\item An \texttt{array} is a comma-separated list of values, which can be any of the six types. An \texttt{array} begins with \texttt{``[''} and ends with \texttt{``]''}.
\end{itemize}

\myfig{}~\ref{fig:json-llmon} shows an example JSON object and how  it can be expressed in LLMON. 
The JSON object contains a dictionary (`Trip`) with two key/value pairs. The first pair is a simple key/value, where the value is a string (`Business`). The second pair has a value that is a list of strings (`New York`, `Tokyo`, `Cairo`).

The conversion to LLMON leverages LLMON's instances naming feature and its explicit nesting feature. 
Rather than using nesting, it creates blocks for each data value (\stt{Business}, 
\stt{New York}, \stt{Tokyo}, and \stt{Cairo}). It tags these blocks using the keys in the nesting from the JSON combined with the \stt{<|.|>} special token (\stt{Trip<|.|>Purpose} or \stt{Trip<|.|>Cities}).
Finally, it encodes each list element with its id (\stt{Trip<|.|>Cities<|:|>0},
\stt{Trip<|.|>Cities<|:|>1},
\stt{Trip<|.|>Cities<|:|>2}),
so it can be referenced by other text as needed.

Although the LLMON version has apparent redundancy in the repeating the outer nesting tag names, we believe this will help LLMs learn the tag semantics for each block.

\ignore{
The JSON object contains two key/value pairs. The first pair is a simple key/value, where the value is a string (\texttt{"Planned Trips"}). The second pair has a value that is a list of strings (\texttt{``New York'', ``Tokyo'', ``Egypt''}).
}

\begin{figure}
\hspace{4.5em}
\textbf{JSON} 
\hspace{15em} 
\textbf{LLMON}\\
\hspace{1em}
\fbox{
\begin{footnotesize}
\begin{tabular}{l}
\{ \\
 \hspace{1em} "Trip": \{ \\
 \hspace{2em}  "Purpose": "Business", \\
 \hspace{2em} "Cities": [ \\
 \hspace{3em}  "New York", \\
 \hspace{3em}  "Tokyo", \\
 \hspace{3em}  "Cairo" \\
 \hspace{2em}       ] \\
 \hspace{1em}   \} \\
\}
\end{tabular}
\end{footnotesize}
}
\hfill
\fbox{
\begin{footnotesize}
\begin{tabular}{l}
<|open|>Trip<|.|>Purpose<|close|>Business<|open\_end|> \\
<|open|>Trip<|.|>Cities<|:|>0<|close|>New York<|open\_end|> \\ 
<|open|>Trip<|.|>Cities<|:|>1<|close|>Tokyo<|open\_end|> \\
<|open|>Trip<|.|>Cities<|:|>2<|close|>Cairo<|open\_end|>
\end{tabular}
\end{footnotesize}
}
\hspace{1em}
\caption{Example JSON and equivalent LLMON. Indentation and spacing are not significant.}
\label{fig:json-llmon}
\Description[]{}
\end{figure}


\subsection{LLMON Converters}
We have created the following converters:
\begin{itemize}
    \item JSON to \mrllmon{}
    \item \mrllmon{} to JSON 
    \item \short{} to \mrllmon{}
    \item \mrllmon{} to \short{}
\end{itemize}
These straightforward converters leverage the grammar from \myfig{}~\ref{fig:llmon-grammar} and libraries for JSON parsing and serialization. 

The \short{} to \mrllmon{} converter provide two features that help to address two design requirements from Section~\ref{sec-properties}:
\begin{itemize}
    \item Escape sequences from \short{} (using ``\tb'') are expanded into the tokens provided in Table~\ref{tab-llmon-special-tokens}.
    \item Any non-flattened nesting tag names are converted to their flattened nesting tags. This allows a human to express nested structures succinctly, but still have the flattened tags being used when passed to LLMs. \myfigs{}~\ref{fig:email-shorthand} and ~\ref{fig:email-mrllmon} in the Appendix provide examples of the succinct \short{} syntax and how it is flattened in \mrllmon{} after conversion.
\end{itemize}

The \short{} to \mrllmon{} converter also supports a comment facility that is expressed
in the \short{} syntax as \stt{\tb comment\tb} ... \stt{/comment/}. Comments are removed during conversion to LLMON.

\section{Leveraging LLMON Information}
\label{sec-use}

This section describes how LLMON-annotated data can be used during two stages of the AI lifecycle, both with the goal of improving accuracy, safety, and security of the model. These stages are (i) model training, including base pretraining and post-training (e.g., fine-tuning), and (ii) inference, where runtime systems assemble, validate, and manage the context used by the model during generation~\citep{brown2020fewshot,ouyang2022instructgpt}.
We focus on the inference-time components where explicit and structural metadata can be recognized and enforced.  

Before either training or inference can occur, input text must first be converted into \emph{tokens} (IDs) by a tokenizer~\citep{sennrich2016subword,kudo2018sentencepiece}. 
Systems may also define \emph{special tokens}—character sequences treated atomically by the tokenizer—to make structural boundaries unambiguous (e.g., chat roles or markup delimiters)~\citep{devlin2019bert}.
Because \mrllmon{} is designed around a small set of special tokens to delimit tag open/close and connector markers, the model and runtime can reliably identify segment boundaries in the token stream, as described in Section~\ref{sec-mrllmon}.

\mypara{Model training: base training and post-training}
Modern LLM development typically starts with \emph{base training (pretraining)} over large sets of text, optimizing the ability of the model to predict the next token in a sequence~\citep{brown2020fewshot}.
This process broadly teaches the model to understand language, but does not impart any application-level distinctions such as \emph{instruction vs.\ data}---a gap that motivates LLMON's explicit segmentation idea~\citep{ouyang2022instructgpt,greshake2023promptinjection}.
Models are then adapted via \emph{post-training}, which commonly includes: \emph{supervised fine-tuning (SFT)} on instruction-response pairs, preference-based alignment methods (e.g., RLHF and DPO-style approaches), and parameter-efficient tuning such as \emph{Low-Rank Adaptation (LoRA)}, which specialize behavior by training a small number of additional parameters rather than updating all weights~\citep{ouyang2022instructgpt,christiano2017preferences,rafailov2023dpo,hu2022lora,houlsby2019adapters}.
A critical point for this paper is that training defines the model's learned interface, often through repeating specific patterns~\citep{wei2022zeroshot}.
If training data consistently presents boundaries and labels for segments, the model can learn to behave more reliably around those boundaries (e.g., treating data parts as nonexecutable content)~\citep{greshake2023promptinjection}.

\mypara{Inference: context management, safety control, and memory efficiency}
During inference, the model generates responses conditioned on the runtime context.
In modern LLM systems, inference is rarely a single forward pass over a static prompt or forward pass; instead, it typically forms a pipeline that performs context construction, validation, and system-level optimizations that jointly affect model behavior and computational cost.
Context management selects and organizes inputs to maximize utility under length constraints.
Representative techniques include prompt optimization/templating, retrieval-augmented generation, or context compression to reduce redundancy and distraction while preserving task-critical evidence. 
Safety control is often implemented via input validation and transformation before generation.
Rather than relying solely on the model to infer potentially adversarial content, such runtime safeguards can explicitly check and rewrite the inputs.
This includes instruction isolation, policy checking, and selective exclusion of untrusted content to mitigate prompt injection and related input-manipulation risks~\citep{greshake2023promptinjection}.
These measures provide application-level guarantees even when the model was not trained to reliably recognize all attack or harmful patterns.
Finally, memory efficiency is critical for scalable deployment.
Inference implementations typically employ key-value (KV) cache assignment, memory sharing across requests, and dynamic batching to reduce redundant computation.
These techniques allow shared or repeated prompt prefixes to be processed once and reused across decoding steps or across requests, improving throughput and latency as context lengths increase.
Taken together, these inference-time optimizations emphasize that an LLM system's behavior is determined not only by learned model parameters but also by the runtime pipeline that constructs and constrains the context. 
LLMON complements these optimizations by providing explicit, well-formed tags that expose segment boundaries to the inference stage.
These data tags enable deterministic preprocessing, more principled context management, and clearer separation among distinct information sources (e.g., intended or unintended content, effective or redundant data), thereby improving reliability and efficiency at inference time.

The following subsections describe how LLMON structure can be leveraged during model training and inference implementations, and how these mechanisms enable more reliable instruction selection, data isolation, and structured execution.


\subsection{Leveraging LLMON Information During Model Training}
\label{sec-use-training}

LLMON introduces explicit structural signals into the training data that are difficult to encode reliably using plain text prompts alone: instruction-data separation, identifier-based reference, and explicit execution intent.
In standard \textit{SFT}, the model is trained over a flat token sequence, and any instruction-like text appearing anywhere in the context may be learned as potentially executable.
LLMON alters this by embedding structural boundaries directly into the token stream (via \mrllmon{}), enabling the model to learn which spans represent \emph{control}, which should be treated as \emph{data}, and how to bind execution to specific referenced instances.


\mypara{What training learns from LLMON}
LLMON-annotated examples provide three key signals.
First, \emph{role separation}, where tokens inside ``\stt{instr}'' spans represent control, while tokens inside ``\stt{data}'' spans represent contextual payload.
Second, \emph{boundary clarity}, where dedicated open/close delimiters make span extents explicit even in long or composite prompts.
Third, \emph{referential binding}, where instance identifiers allow execution to reference specific instruction or data objects rather than relying on positional cues such (e.g., ``the above text'').
Together, these signals encourage the model to condition behavior on explicitly selected instructions while treating non-selected spans as non-authoritative context.


\mypara{How this changes post-training}
The optimization objective remains standard causal language modeling, we update only the sequence serialization.
Instead of learning from loosely structured narrative prompts, the model is trained on structured interface artifacts containing multiple candidate instructions, explicit data objects, and an ``\stt{exec}'' span selecting what should run.
This shifts instruction following from implicit formatting conventions toward explicit token-level semantics.


\mypara{Why special tokens matter}
\mrllmon{} uses a minimal set of dedicated special tokens that the tokenizer treats atomically.
This ensures that structural delimiters are consistently represented in the token stream and are not fragmented or merged during tokenization.
As a result, boundaries become stable, learnable elements across training runs and model families, providing a persistent structural ``control plane'' independent of the natural-language content inside spans.


\mypara{Compatibility and incremental adoption}
LLMON operates entirely at the interface layer.
It requires no model architectural modification and is compatible with standard post-training methods, including full fine-tuning and parameter-efficient approaches such as LoRA~\citep{hu2022lora}.
LLMON-annotated examples can be mixed with conventional instruction-tuning data, enabling gradual adoption while preserving general instruction-following capability.


\subsection{Creating LLMON training data}
\label{sec-use-data}

Operationalizing LLMON requires constructing training corpora in which structural roles and execution intent are explicitly encoded in the token stream.
Rather than modifying model architectures or objectives, we transform existing post-training and instruction-tuning datasets into structured interface artifacts expressed in \mrllmon{}.

Starting from conventional \textit{SFT} datasets, such as Cleaned Alpaca~\citep{taori2023stanford} and Dolly~\citep{conover2023free}, as well as broader public instruction-tuning corpora~\citep{ouyang2022instructgpt}, each example is serialized by placing the instruction inside an ``\stt{instr}'' span and the optional input inside a ``\stt{data}'' span, followed by an explicit ``\stt{exec}'' span that binds the two together.
The target output remains unchanged, preserving original task semantics while making control and data boundaries explicit.
An example is shown in \myfig{}~\ref{fig:llmon-intro-mrllmon}.

To encourage structured execution behavior under competing signals, we additionally construct distractor-infused LLMON variants.
These examples include one ``in-focus'' instruction together with optional non-selected instruction spans and unrelated data spans.
The ``\stt{exec}'' span explicitly binds execution to the correct instance, teaching the model to rely on identifier-based selection rather than positional or heuristic cues.
These structured variants are designed to train focus and referential binding in multi-instruction settings.
One such example is shown in \myfig{}~\ref{fig:exec-noargs-mrllmon}.

The final training mixtures combine (i) LLMON-wrapped public instruction data, which preserves original tasks while adding explicit structure, and (ii) distractor-infused LLMON variants that stress instruction-data separation.
Across datasets, this process scales to millions of structured examples and billions of tokens (Section~\ref{sec-eval}).
Significantly, the training pipeline remains otherwise unchanged: only prompt serialization and the addition of a small set of special tokens are required.


\subsection{Leveraging LLMON Information During Model Inference}
\label{sec-use-inference}

Here we discuss how the explicit notations provided by LLMON could be leveraged to control the model behavior during inference. Designated LLMON tags can be used to trigger specific mechanisms such as ``thinking'' mode, invoking external tools, or dynamically enabling parameter-efficient adaptations such as Activated LoRA~\citep{greenewald2025activated}. A different form of model's behavior control can take advantage of the segment boundaries from LLMON. A fundamental building block in the LLM transformer architecture is the attention mechanism, which evolves internal token representations by adding to them weighted averages of representations of other tokens. This mechanism enables long distance correspondence between different parts of the sequence which is key to this architecture's success. However, allowing all tokens in the sequence that precedes it is not always beneficial as it may cause hallucinations, create security risks, and degrade efficiency. A mitigation strategy is to mask parts of the context, deemed irrelevant or risky for the current stage of computation, during the attention calculation. Here we can use LLMON tags and segment boundaries to determine which segments should be masked and the exact positions of the tokens to be masked. 

Two important use cases are instruction selection and prompt rejection. For instruction selection, multiple candidate instructions can coexist in the same input sequence. A query determines which instruction should be invoked at a given stage of the calculation (see example in \myfig{}~\ref{fig:exec-noargs-mrllmon}). By masking the distracting instructions, we can guarantee that during generation, the only tokens which are attended to are from the instruction to be invoked, ensuring the correct behavior. Similarly, for prompt rejection, untrusted or policy-violating content can be systematically excluded from the response's attention, guaranteeing that it does not influence the forward generation. Because these guarantees would be derived from explicit boundaries rather than probabilistically learned boundary detection, they remain stable under prompt variation, domain shift, and adversarial formatting. Compared with training-based or heuristic ways to obtain separation, such an approach would yield more reliable parsing and reduce unintended cross-region interference.

\section{Initial Evidence of Value}
\label{sec-eval}

Although a full evaluation of the use of LLMON information across tasks and model families is beyond the scope of this paper, we present some initial empirical evidence of LLMON's efficacy for the two use cases described in Section~\ref{sec-use}: model training and model inferencing.

\subsection{Leveraging LLMON Information During Model Training} \label{sec-eval-training}
This section examines whether explicit structural annotation during model training improves robustness and execution control without degrading general instruction-following performance.


\mypara{Research questions}
This evaluation investigates whether explicit instruction-data separation improves robustness and execution control.
More concretely, we consider the following questions:
\begin{description}
    \item [RQ1:] Does explicit instruction-data separation improve robustness to distractor or competing instructions?
    \item [RQ2:] Can models learn reliable referential binding through identifier-based execution calls?
    \item [RQ3:] Does enforcing structured execution degrade standard benchmark performance?
    \item [RQ4:] How do these effects vary across post-training regimes (full fine-tuning vs.\ LoRA)?
\end{description}

These questions directly correspond to the failure modes identified in Section~\ref{sec-llmon} and to the structured training methodology described in Section~\ref{sec-use-data}.

\mypara{Backbone models}
We evaluate two backbone models representing different architectural families and training lineages: Granite-4.0-Micro-Base~\citep{granite2024granite} (3B parameters) and Qwen2.5-3B~\citep{qwen2025qwen25technicalreport}.
These models serve as representative mid-scale LLMs with competitive baseline performance.
For each backbone, we report results for (i) the base model without post-training, (ii) conventional post-training using chat-template instruction data, and (iii) post-training with LLMON-structured corpora under both full fine-tuning and LoRA adaptation.

\mypara{Training configuration and scale}
We use the structure-aware corpora described in Section~\ref{sec-use-data}, comprising (i) LLMON-wrapped public post-training and instruction-tuning data, and (ii) distractor-infused LLMON variants designed to train identifier-based focus under multi-instruction conditions.
For the baseline post-training runs, we use the same underlying instruction-tuning datasets serialized using conventional chat-template formatting.
Across large-scale public single-turn \textit{SFT} data, Alpaca, and Dolly, the resulting mixture contains roughly \textbf{3.4 million structured examples} totaling \textbf{2.9 billion tokens}.
To study scaling effects, we train models on progressive subsets of this corpus (241M, 436M, 616M tokens) as well as the full 2.9B-token mixture.
Post-training is performed using standard \textit{SFT} under both full fine-tuning and parameter-efficient LoRA adaptation. No architectural modifications are introduced.

\mypara{Evaluation protocol}
We introduce a dedicated \emph{Distractor} benchmark: 100 manually curated Alpaca instances in multi-instruction form with one instruction selected via ``\stt{exec}''.
An LLM-as-a-Judge (LLMaJ) scores each response, measuring identifier-based binding and robustness to distractors.
The final score is the average across all instances.
We measure robustness and execution control (RQ1--RQ2) using the Distractor benchmark, which evaluates whether a model executes the instruction explicitly referenced by the ``\stt{exec}'' in multi-instruction contexts.
Performance reflects correct identifier-based instruction binding.
General capability retention (RQ3) is evaluated using standard single-instruction benchmarks without distractors, including MMLU~\citep{hendrycks2021measuringmathematicalproblemsolving}, GSM8K~\citep{cobbe2021trainingverifierssolvemath}, and IFEval~\citep{zhou2023instructionfollowingevaluationlargelanguage}.

\mypara{Results}
Table~\ref{tab-llmon-ft-results} compares base models, conventional post-training baselines using chat-template instruction data, and LLMON-structured post-training.
The Distractor column demonstrates that structured post-training with LLMON substantially improves execution control for both backbones.
Both base models and conventionally post-trained chat-template baselines (full fine-tuned) exhibit near-zero (0.00 and 0.40) accuracy on the Distractor benchmark, indicating weak identifier-based binding and frequent execution of positionally salient or distractor instructions.
After LLMON-structured fine-tuning, execution accuracy rises dramatically to above 83\% across token scales, confirming that explicit instruction-data separation and ``\stt{exec}''-based binding can be reliably learned.
In contrast, models post-trained using conventional chat-template formatting show little improvement and often degrade performance on the Distractor benchmark, indicating that the gains arise from structured supervision rather than from additional post-training alone.

\begin{table}[ht]
\begin{center}
\begin{footnotesize}
  \caption{Comparison of base models, conventional post-training baselines (chat-template instruction tuning), and LLMON-structured post-training.
  MMLU, GSM8K, and IFEval are reported in zero-shot accuracy, and ``Distractor'' denotes accuracy on our structured multi-instruction Distractor benchmark. Token counts indicate the amount of post-training data used.}
  \label{tab-llmon-ft-results}
  \begin{tabular}{lc|rrrr}
    \toprule
    Model          & Tokens & MMLU & GSM8K & IFEval & Distractor \\
    \midrule
    \multicolumn{6}{l}{\textit{\textbf{Base (no post-training)}}} \\
    \midrule
    Granite-4.0-Micro-Base  & --   & 61.52 & 0.00  & 38.78 &  4.40 \\
    Qwen-2.5-3B      & --   & 65.09 & 0.15  & 27.13 & 27.80 \\
    \midrule
    \multicolumn{6}{l}{\textit{\textbf{Full fine-tuning baseline (post-training with chat-template data)}}} \\
    \midrule
    Granite-4.0-Micro-Base  & 2.9B & 39.95 & 0.61  & 72.07 &  0.00 \\
    Qwen-2.5-3B      & 2.9B & 48.51 & 1.13  & 74.05 &  0.40 \\
    \midrule
    \multicolumn{6}{l}{\textit{\textbf{Full fine-tuning (post-training with LLMON-structured data)}}} \\
    \midrule
    \multirow{4}{*}{Granite-4.0-Micro-Base}
      & 241M & 36.69 & 26.31 & 54.75 & 87.80 \\
      & 436M & 36.38 & 13.12 & 57.95 & 87.40 \\
      & 616M & 41.95 & 15.84 & 65.82 & 85.80 \\
      & 2.9B & 43.35 & 20.39 & 68.14 & 84.00 \\
    \midrule
    \multirow{4}{*}{Qwen-2.5-3B}
      & 241M & 48.35 & 23.65 & 58.86 & 88.00 \\
      & 436M & 48.01 & 24.18 & 62.40 & 87.60 \\
      & 616M & 47.92 & 12.21 & 65.38 & 86.80 \\
      & 2.9B & 47.26 & 16.83 & 71.96 & 83.40 \\
    \midrule
    \multicolumn{6}{l}{\textit{\textbf{LoRA baseline (post-training with chat-template data)}}} \\
    \midrule
    Granite-4.0-Micro-Base  & 2.9B & 58.53 &  7.43 & 67.08 & 19.40 \\
    Qwen-2.5-3B      & 2.9B & 63.55 & 14.93 & 47.26 &  2.40 \\
    \midrule
    \multicolumn{6}{l}{\textbf{\textit{LoRA (post-training with LLMON-structured data)}}} \\
    \midrule
    \multirow{4}{*}{Granite-4.0-Micro-Base}
      & 241M & 53.72 & 34.87 & 54.07 & 74.40 \\
      & 436M & 51.32 & 20.24 & 50.58 & 75.60 \\
      & 616M & 56.11 & 7.05  & 56.22 & 74.20 \\
      & 2.9B & 54.97 & 15.77 & 60.35 & 75.00 \\
    \midrule
    \multirow{4}{*}{Qwen-2.5-3B}
      & 241M & 49.57 & 4.47  & 33.07 & 72.20 \\
      & 436M & 62.63 & 7.43  & 38.80 & 72.40 \\
      & 616M & 48.31 & 24.03 & 44.57 & 69.80 \\
      & 2.9B & 55.62 & 20.24 & 48.20 & 70.80 \\
    \bottomrule
  \end{tabular}
  \end{footnotesize}
\end{center}
\end{table}

On fully fine-tuned models, scaling effects show that increasing token budgets generally stabilizes structured behavior and improves performance on instruction-following benchmarks such as IFEval, though gains are not strictly monotonic across all tasks.
Significantly, LLMON-structured supervision does not catastrophically degrade general capability: while some benchmarks fluctuate relative to base performance, models retain competitive accuracy on MMLU, GSM8K, and IFEval.

LoRA adaptation also yields large improvements in LLMON-structured execution relative to base models (typically around 70\% compared to near-zero baseline performance), demonstrating that structural behavior can be acquired in a parameter-efficient regime.
However, LoRA consistently underperforms full fine-tuning on the Distractor benchmark, particularly at larger token scales, suggesting that deeper weight updates better internalize execution semantics.
Interestingly, LoRA sometimes achieves stronger performance on knowledge-oriented benchmarks such as MMLU, whereas IFEval tends to benefit more consistently from full fine-tuning, and GSM8K shows mixed results across models.
Across both backbones, full fine-tuning consistently achieves the highest Distractor benchmark scores, reinforcing the central claim that LLMON-structured supervision improves control and multi-instruction robustness rather than merely fitting surface patterns.

These results directly address RQ1--RQ4.
The large gains on the Distractor benchmark
(an average improvement of 86.2 and 62.2 percentage points for full-fine tuning and LoRA, respectively)
demonstrate improved robustness to distractor instructions (RQ1) and reliable identifier-based execution binding (RQ2).
The absence of catastrophic degradation on standard single-instruction benchmarks (MMLU, GSM8K, IFEval) supports that structured supervision preserves general capability (RQ3).
Finally, comparing full fine-tuning and LoRA shows that structured execution behavior can be learned under both regimes, with full fine-tuning yielding stronger and more stable execution accuracy at scale (RQ4).


\subsection{Leveraging LLMON Information During Model Inference} \label{sec-eval-inference}
This section examines how the explicit segment boundaries from LLMON can improve robustness and execution control during model inference.

\begin{table}[t]
\begin{center}
\begin{footnotesize}
\caption{Comparison of models applying constraint masking with LLMON at inference on multi-instruction Distractor benchmark.}
\begin{tabular}{llc}
\toprule
Model & Method & Distractor \\
\midrule
\multirow{2}{*}{Qwen2.5-3B-Instruct}
 & baseline & 43.20 \\
 & constraint mask with LLMON & \textbf{69.00} \\
\midrule
\multirow{2}{*}{Granite-3.3-8B-Instruct}
 & baseline & 41.60 \\
 & constraint mask with LLMON & \textbf{72.00} \\
\midrule
\multirow{2}{*}{Granite-4.0-Micro}
 & baseline & 40.80 \\
 & constraint mask with LLMON & \textbf{72.40} \\
\bottomrule
\end{tabular}
\label{tab-inference-distractor-results}
\end{footnotesize}
\end{center}
\end{table}

\mypara{Research questions}
This evaluation investigates whether the presence of explicit instruction-data separation during model inferencing improves robustness and execution control.
Here we do not assume any LLMON post-training, but instead rely solely on inference-time techniques.
Specifically, we apply boundary-constrained masking (described in Section~\ref{sec-use-inference}), which parses LLMON annotations to identify boundaries and applies structured masking strategies that regulate generative computation according to explicit delimiters.
This enables deterministic instruction selection and data isolation at inference time. We consider these questions:
\begin{description}
    \item[RQ5:] Does explicit boundaries influence the model behaviors at \emph{model inference}?
    \item[RQ6:] Does instruction-data separation \emph{at model inference} without training improve robustness to distractor or competing instructions?
\end{description}

\mypara{Models}
For the inference evaluations, we select three models: Qwen2.5-3B-Instruct, Granite-3.3-8B-Instruct, and Granite-4.0-Micro. We use instruction models to evaluate inference effectiveness, because their instruction capabilities are ready for the inference tests and provide a consistent capability baseline. This allows us to more clearly isolate and compare the effects of constrained masking with LLMON. In contrast, base models (used in Section~\ref{sec-eval-training}) are designed for post-training only and are less suitable for this evaluation, as their weaker instruction behavior may introduce additional variability that could confound the results.

\mypara{Evaluation protocol}
We use the same Distractor benchmark and metric discussed in Section~\ref{sec-eval-training}.

\mypara{Results}
Table~\ref{tab-inference-distractor-results} presents the results. For each model, we report baseline performance on the Distractor benchmark without leveraging any LLMON information, as well as results obtained when LLMON information is applied through constrained masking at inference time. In both settings, LLMON information is not used during training.

The results in Table~\ref{tab-inference-distractor-results} show it is feasible to incorporate explicit boundaries into the constrained token masking as reference during inference time (RQ5). The performance comparisons show that applying constrained masking with LLMON boundaries at inference time consistently improves robustness to distractor instructions across all models (RQ6). In the baseline setting, performance is similar across well-trained instruction models, ranging from 40.8 to 43.2, indicating that different model sizes exhibit comparable vulnerability to distractor instructions. When LLMON information and constrained masking are applied during inference, performance increases
with values of 69.0 to 72.4, yielding improvements of 25.8 to 31.6 points, for an average of 29.3. Notably, these gains are consistent across models, suggesting that the benefits are sustainable and do not depend on model scale. Overall, these findings show an example of leveraging LLMON information at inference time to mitigate distractor instruction interference without retraining.

Both training-time (Section~\ref{sec-eval-training}) and inference-time (Section~\ref{sec-eval-inference}) evaluations that leverage LLMON information achieved significant improvements on the Distractor benchmark across multiple models. Training-time results improved by 
an average of 74.2 percentage points.
Inference time results improved by an average of 29.3 percentage points.
Together, these findings demonstrate two complementary pathways for enhancing model robustness: incorporating LLMON information during model training or leveraging it at model inference time.

\section{Related Work}
\label{sec-related}
This section discusses related work to the notion of an LLM-native markup language.

\mypara{Chat templates}
LLMs are commonly trained to follow chat templates that serialize multiple interactions into a single sequence of tokens, delineating roles such as system, user, and assistant. In practice, these delimiters range from plain-text prefixes (``\stt{User:}'') to dedicated special tokens that are treated atomically by the tokenizer~\citep{kudo2018sentencepiece,hf_chat_templating,mistral_chat_templates,ibm_granite_prompt_guide_4}. Even with atomic role markers, a transformer does not automatically enforce role separation or instruction-vs-data boundaries beyond the statistical behavior learned during training. As a result, when untrusted content is included in the same context as instructions, models can be coerced into interpreting data as directives, enabling prompt-injection and related attacks~\citep{greshake2023promptinjection}. LLMON, in contrast, introduces an explicit span layer with delimiter tokens that mark the start and end of spans. These defined boundaries enable the model to be conditioned on them consistently and can be enforced by downstream runtimes. The effect is a separation of representation from enforcement, rather than relying on prompt conventions alone.

\mypara{What role does JSON play}
JavaScript Object Notation (JSON) is ubiquitous, broadly supported, and tool-friendly~\citep{rfc8259}. It is, however, insufficient as an LLM-native interface because it lacks token-level and span-level control~\citep{vaswani2017attention}.
JSON's schema-centric model and quoting/escaping rules are valuable for deterministic parsers, yet they add verbosity and rely on arbitrary wrapper fields to specify instruction/data separation, annotations, and unique references~\citep{rfc8259}. As JSON delimiters are not reserved special tokens, they are destined to be tokenized on sub-word boundaries~\citep{kudo2018sentencepiece,devlin2019bert}. With no tokenizer/runtime hooks to reliably isolate untrusted spans, JSON is prone to methods of indirect prompt injection~\citep{greshake2023promptinjection}. In contrast, LLMON provides explicit span identifiers and a compact form with special tokens as delimiters. This enables structure-aware decoding/enforcement compatible with self-attention~\citep{vaswani2017attention}, KV-cache-based optimization~\citep{kwon2023pagedattention}, and parameter-efficient adaptation (e.g. LoRA)~\citep{hu2022lora}.

\mypara{Prompt Orchestration Markup Language}
POML~\citep{zhang2025promptorchestrationmarkuplanguage,poml-github} is a markup language focusing on adding structure to prompts as well as improving maintainability and versatility. It includes a VSCode extension to ease development and has an SDK to help with integration. It has a rich collection of built-in semantic components (tags). 
Although similar in that LLMON can also be used to add semantic information to prompts, the goals of the efforts are complementary. LLMON is trying to improve the performance and security of LLMs, whereas POML is trying to increase the productivity of the prompt engineer. This means LLMON can, in fact, express POML (or POML-like) semantic components as tags.

\mypara{Prompt engineering frameworks}
More broadly, researchers have proposed frameworks and abstractions  
to help developers create and manage prompts (e.g.,\citep{prompt-frameworks-survey-2026,langchain,guidance,llamaindex,
dspy-2023,LMQL-2023,dspy-2023,autogen-2024,MTP-2025}), sometimes providing
prompt generation techniques~\citep{MTP-2025} or  the expression of parallelism 
in the execution of tool calls~\citep{opal-2025}.
This area of research is complementary to LLMON.
Its goal is to increase the productivity of the prompt engineer or the runtime system performance of tool calling, whereas LLMON is focusing on model performance (accuracy) and security. 

\mypara{The role of Constrained Decoding}
Constrained decoding~\citep{picard-2021} is a LLM-independent inference-time technique that
ensures LLMs generate tokens that satisfy user-specified
constraints, such as regular expressions~\citep{LMQL-2023,willard2023efficientguidedgenerationlarge,guidance}, templates, such as a JSON schema, or parser-guided approaches~\citep{picard-2021}. 
Constrained decoding acts like a validating filter on LLM
generation. It does not allow noncompliant tokens to be generated.
Our approach differs in that we are 
enabling the specification of structure and semantics that can be used for training LLMs and model inference enforcement. The two approaches are complementary and can be used together.

\mypara{Specialized models for generating SQL output}
Prior work has explored restricting auto-regressive decoding~\citep{yin-neubig-2018-tranx,lin2019grammarbasedneuraltexttosqlgeneration, wang-etal-2020-rat} or semi-auto-regressive~\citep{rubin-berant-2021-smbop} to 
token sequences that correctly parse to SQL
abstract syntax trees. Although effective, these approaches
require a custom vocabulary of special tokens, a custom model architecture, or both.
In contrast, LLMON targets the \emph{interface} rather than a SQL-specific semantic parser. It uses explicit span identifiers and delimiter markers intended as reserved special tokens~\citep{devlin2019bert}.
Because these token-level boundaries are preserved atomically during tokenization~\citep{kudo2018sentencepiece}, they make structure-aware decoding and enforcement practical at runtime, without modifying the transformer itself~\citep{vaswani2017attention}.
This also aligns with KV-cache-based serving optimizations~\citep{kwon2023pagedattention} and supports parameter-efficient adaptation (e.g., LoRA)~\citep{hu2022lora}, allowing the same interface abstractions to generalize beyond SQL to other structured-output tasks.

\section{Discussion} \label{sec-disc}
This section discusses broader issues that may arise when pursuing an LLM-native markup language.


\subsection{Open Design Questions and Systems Implications}
Several design choices remain open and are particularly relevant to programming-systems audiences.
First, the tradeoff between verbose and parsimonious closing tags may affect both learnability and ``toolability''.
Second, the choice and initialization of special tokens may influence convergence and stability; one plausible approach is initializing new tokens from semantically related embeddings rather than random initialization.
Third, LLMON opens the door to richer static analyses and developer tooling (formatters, linters, type-like checks over required spans) and to runtime enforcement mechanisms that operate at token/span granularity (e.g., KV-cache and attention masking).
We view these as natural extensions of treating the LLM interface as a programmable artifact rather than an unstructured string.


\subsection{Practical Deployment Considerations}
Some practical concerns warrant further investigation. First, \emph{tokenization efficiency}: whether the six special tokens impact convergence across different tokenizer families (e.g., BPE~\citep{sennrich2016neural}, SentencePiece~\citep{kudo2018sentencepiece}, WordPiece~\citep{devlin2019bert}). Second, \emph{inference overhead}: Boundary-constrained masking requires parsing LLMON spans; computational cost and memory impact on high-throughput serving remain uncharacterized.


\subsection{Future Evaluations}

The goal of this paper is to focus on the opportunity for adding structure and semantics to LLM interactions, define a markup language, and present preliminary evidence that the idea warrants further evaluation and research.
Although the experiments in Section~\ref{sec-eval} demonstrate that models can i) learn structured execution patterns from LLMON-annotated data and ii) inference techniques can leverage LLMON-annotated data to improve robustness, a comprehensive evaluation of the design space remains an important direction for future work. This includes evaluating the value of combining these two complimentary uses of LLMON annotations.

One natural direction is to evaluate the approach across a broader range of model sizes and architectural families.
The experiments presented in this paper focus on representative mid-scale models, but it remains important to understand how LLMON behaves across larger models.
Larger models may already internalize certain structural conventions implicitly, while smaller models may benefit more strongly from explicit supervision signals.
Similarly, evaluating models from additional families could help determine whether the benefits of LLMON arise primarily from the training signal itself or from interactions with specific architectural or design choices.

Another important avenue concerns the types of capabilities being evaluated.
The current experiments focus primarily on robustness to distractor instructions.
However, many tasks that LLMs perform involve strong structural components (e.g., code generation), where LLMs may benefit particularly from explicit structural markup, and future evaluations could measure whether LLMON improves model reliability on tasks where hierarchical or referential structure plays a central role.

Future work should also explore the role of LLMON-style annotations at different stages of the model training pipeline.
The experiments in this paper apply LLMON during post-training (instruction tuning), but it is plausible that introducing structured annotations during pretraining could further strengthen the model's ability to recognize and reason about structured spans.
Understanding how LLMON interacts with pretraining objectives, instruction tuning, and alignment methods remains an open question.

Another potential research direction involves the interaction between LLMON and tokenization strategies.
LLMON relies on a small set of special tokens to delimit structural boundaries and encode relationships such as execution bindings.
Investigating how different tokenizers represent these markers, and whether alternative tokenization schemes improve the learnability of structural annotations, could provide additional insights into how models internalize structured interfaces.
A related question concerns how the embeddings of these special tokens should be initialized during training, since they represent structural delimiters rather than natural-language content.

Finally, further work is needed to understand how LLMON interacts with inference-time system techniques such as constrained decoding and tool orchestration.
LLMON annotations could also serve as structured signals for activating specialized model behaviors, such as triggering chain-of-thought or ``thinking'' modes, invoking external tools, or dynamically enabling parameter-efficient adaptations such as Activated LoRA~\citep{greenewald2025activated}.
Because LLMON explicitly identifies spans, such as instructions, data artifacts, and execution bindings, it creates opportunities for runtime systems to treat these segments differently, potentially enabling stronger guarantees about execution behavior or information flow.
Exploring these possibilities could help bridge the gap between prompt-based interaction patterns and more structured, program-like interfaces for language models.


\subsection{Approaches for Creating LLMON Training Data}

As discussed in Section~\ref{sec-use-data}, LLMON can be used to express concepts and structure during the training of an LLM.
When existing data is already structured, techniques such as the ones in Section~\ref{sec-use-data} can be used to \textit{LLMONize} it, i.e., convert it to LLMON form by introducing explicit annotations for instructions, inputs, and execution bindings while preserving the original task semantics.
Many existing datasets already contain implicit structure (e.g., instruction-response pairs, question-answer datasets, or structured formats such as JSON) which can often be converted to LLMON automatically through deterministic transformations.

When no explicit structure is present in the training data, additional methods may be required to infer it.
One approach is human annotation, where domain experts identify instructions, data spans, and other relevant concepts within the text.
Another approach is to employ an LLM-based annotator or judge model that analyzes raw text and proposes candidate LLMON annotations, which can then be validated or refined.
Such automated approaches may enable large-scale generation of structured training data, though further experimentation is needed to determine their reliability and cost effectiveness.

Regardless of the approach used, \textit{LLMONized} training data can be added incrementally.
Because LLMON operates primarily at the interface layer, structured examples can be mixed with conventional training data without requiring architectural changes to the model or training pipeline.
This enables gradual adoption in which users can begin by annotating a subset of data where structure is most valuable, while continuing to leverage existing corpora in their original form.


\subsection{Governing the User Tag Namespace}

LLMON introduces the concept of user-defined tags, allowing flexibility in the concepts that an LLM can receive. However, with this flexibility comes a governance problem: who is responsible for managing this global namespace for such tags.
For example, if a base model is trained with data that uses ``\stt{poem}'' to suggest a certain semantics, when is it appropriate to use that tag for post-training? This raises questions about \textit{name collisions} (two distinct concepts using the same tag name) and \textit{aliases} (two different tags being used to represent the same concept). Both would likely lead to reduction in model performance.
One option is to have the training organization govern the namespace to ensure it is used consistently and also to annotate the training data (with tools) as needed. Although appealing, this means that other organizations that want to perform post-training should either use  the approved annotation toolchain or not include any annotations (to avoid namespace conflicts). It is not clear how feasible this constraint will be in practice.


\subsection{A Trend Toward Grammar-like Ideas?}

Normally LLMs are not given any grammar, just the training data.
Although tokenization captures some very frequent ``small'' patterns, by collapsing them into tokens, bigger structures are not explicitly captured.
The LLMON approach is also not specifying a grammar, but it is providing information about structure beyond tokenization.
It is a step on the ``no grammar'' $\rightarrow$ ``grammar'' spectrum. 
Likewise, constrained decoding uses a simplified grammar during generation.
Thus, there seems to be the beginnings of a trend towards adopting grammar-like ideas into LLM generation/training.
The types of LLMON go further than grammars by adding the traditional semantic analysis (aka ``type checking'') component of a compiler.


\subsection{Analogies to Computer Architecture ISAs}

Computer architectures provide the ISA abstraction layer that defines the primitive instructions that the system can execute.
This includes instructions such as ``\stt{Add}'', ``\stt{Sub}'', ``\stt{Compare}'', ``\stt{Branch on Equal}'', etc.
The semantics of these instructions is well specified.
Software is written to use this ISA via a compiler and the ISA is implemented by the manufacturer.

One may wonder how the tag ``\stt{exec}'' used in Section~\ref{sec-examples} compares to an ISA instruction in that both seem to be an abstraction of a primitive instruction.
In fact, an ``\stt{exec}'' instruction is quite similar to a ``\stt{Call}'' instruction that has the call target as the operand along with its parameters.

Here are some comparisons:
\begin{itemize}
    \item ISA instructions have well-defined, guaranteed semantics, an \stt{Add} of two integers will be the same as the mathematical add operation. A LLMON instruction can have this same guarantee using the inference implementation techniques from Section~\ref{sec-use-inference}. If only the training techniques from Section~\ref{sec-use-training} are used, the semantics would not be guaranteed.
    \item ISA instructions are fixed in number and defined at ISA design time. A new LLMON instruction can be created by the user via training data to define its semantics.
\end{itemize}
It will be interesting to see the efficacy of other types of ISA instructions in the LLMON context, such as comparisons, branches, indirect references, etc.
\section{Conclusions} \label{sec-conc}
This work introduces the potential value of an LLM-native markup language. 
We have enumerated the design requirements for such a language and introduced LLMON as a markup language that satisfies these requirements.
The language can be expressed in human-friendly \short{} syntax which can be converted to the equivalent LLMON syntax that is 
communicated to LLMs.
We then described how this additional metadata can benefit two important use cases: model training and inference implementation.
Through preliminary experiments we show the promise of this approach for these two use cases, resulting in average improvements of 
74.2 and 29.3
percentage points, respectively. We enumerate future directions that include additional evaluations and numerous other research opportunities.

\section*{Acknowledgments}
We would like to thank David Grove, Kush Varshney, Martin Hirzel, and Danish Contractor for feedback on earlier versions of this work.
\bibliographystyle{ACM-Reference-Format}
\bibliography{refs}

\appendix
\newpage

\section{LLMON Examples}
This section contains additional LLMON examples.
\myfig{}~\ref{fig:email-shorthand} shows how the structure and semantics of an email message can be expressed in \short{} syntax.
\myfig{}~\ref{fig:email-mrllmon} shows the equivalent email expressed in \mrllmon{}. 
The \short{} syntax uses familiar nesting structures, which the converter converts to a more LLM-friendly format where implied nesting is explicitly described in the tag.

We also show the \mrllmon{} version of examples from the paper that were expressed in \short{}.
\myfig{}~\ref{fig:exec-1arg-llmon} shows the \mrllmon{} version of \myfig{}~\ref{fig:exec-1arg}.
\myfig{}~\ref{fig:prompt-injection-llmon} shows the \mrllmon{} version of \myfig{}~\ref{fig:prompt-injection-shorthand}.

\begin{figure}
\begin{center}
\fbox{
\begin{footnotesize}
\begin{tabular}{l}
\textbackslash email\textbackslash \\
 \hspace{1em} \textbackslash header\textbackslash \\
 \hspace{2em} \textbackslash from\textbackslash alice@example.com/from/ \\
 \hspace{2em} \textbackslash to\textbackslash bob@example.com/to/ \\
 \hspace{2em} \textbackslash subject\textbackslash Design docs/subject/ \\
 \hspace{2em} \textbackslash smpt/ \\
 \hspace{1em} /header/ \\
 \hspace{1em} \textbackslash body\textbackslash \\
 \hspace{2em} \textbackslash paragraph\textbackslash Please see the attachments./paragraph/ \\
 \hspace{2em} \textbackslash notes\textbackslash We’ll review at 2 PM./notes/ \\
 \hspace{1em} /body/ \\

\hspace{1em} \textbackslash attachments\textbackslash \\
\hspace{2em} \textbackslash attachment:1\textbackslash \\
\hspace{3em} \textbackslash filename\textbackslash design\_spec.pdf/filename/ \\
\hspace{3em} \textbackslash type\textbackslash pdf/type/ \\
\hspace{2em} /attachment:1/ \\
\hspace{2em} \textbackslash attachment:2\textbackslash \\
\hspace{3em} \textbackslash filename\textbackslash design\_spec.pdf/filename/ \\
\hspace{3em} \textbackslash type\textbackslash pdf/type/ \\
\hspace{2em} /attachment:2/ \\
\hspace{1em} /attachments/ \\
/email/ \\
\end{tabular}    
\end{footnotesize}
}
\caption{LLMON email example expressed in \short{}}
\label{fig:email-shorthand}
\Description[]{}
\end{center}
\end{figure}


\begin{figure}
\begin{center}
\fbox{
\begin{footnotesize}
\begin{tabular}{l}
\lmo email\lmc \\
 \hspace{1em} \lmo email<|.|>header\lmc \\
 \hspace{2em} \lmo email<|.|>header<|.|>from\lmc \\
   \hspace{3em}  alice@example.com \\
 \hspace{2em}    \lmoe email<|.|>header<|.|>from\lmc \\
 \hspace{2em} \lmo email<|.|>hearder<|.|>to\lmc \\
 \hspace{3em}    bob@example.com \\
 \hspace{2em}    \lmoe email<|.|>hearder<|.|>to\lmc \\
 \hspace{2em} \lmo email<|.|>hearder<|.|>subject\lmc \\
 \hspace{3em}    Design docs \\
 \hspace{2em}    \lmoe email<|.|>hearder<|.|>subject\lmc \\
 \hspace{2em} \lmo email<|.|>hearder<|.|>smpt\lmc \\
 \hspace{1em} \lmoe email<|.|>header\lmc \\
 \hspace{1em} \lmo email<|.|>body\lmc \\
 \hspace{2em} \lmo email<|.|>body<|.|>paragraph\lmc \\
  \hspace{3em}       Please see the attachments. \\
 \hspace{2em} \lmoe email<|.|>body<|.|>paragraph\lmc \\
 \hspace{2em} \lmo email<|.|>body<|.|>notes\lmc \\
 \hspace{3em} We’ll review at 2 PM. \\
 \hspace{2em} \lmoe email<|.|>body<|.|>notes\lmc \\
 \hspace{1em} \lmoe email<|.|>body\lmc \\

\hspace{1em} \lmo email<|.|>attachments\lmc \\
\hspace{2em} \lmo email<|.|>attachments<|.|>attachment<|:|>1\lmc \\
\hspace{3em} \lmo email<|.|>attachments<|.|>attachment<|:|>1<|.|>filename\lmc design\_spec.pdf \\
\hspace{3em} \lmoe email<|.|>attachments<|.|>attachment<|:|>1<|.|>filename\lmc \\
\hspace{3em} \lmo email<|.|>attachments<|.|>attachment<|:|>1<|.|>type\lmc pdf \\
 \hspace{3em} \lmoe email<|.|>attachments<|.|>attachment<|:|>1<|.|>type\lmc \\
\hspace{2em} \lmoe email<|.|>attachments<|.|>attachment<|:|>1\lmc \\
\hspace{2em} \lmo email<|.|>attachments<|.|>attachment<|:|>2\lmc \\
\hspace{3em} \lmo email<|.|>attachments<|.|>attachment<|:|>2<|.|>filename\lmc design\_spec.pdf \\
\hspace{3em} \lmoe email<|.|>attachments<|.|>attachment<|:|>2<|.|>filename\lmc \\
\hspace{3em} \lmo email<|.|>attachments<|.|>attachment<|:|>2<|.|>type\lmc pdf \\
\hspace{3em} \lmoe email<|.|>attachments<|.|>attachment<|:|>2<|.|>type\lmc \\
\hspace{2em} \lmoe email<|.|>attachments<|.|>attachment<|:|>2\lmc \\
\hspace{1em} \lmoe email<|.|>attachments\lmc \\
\lmoe email\lmc \\
\end{tabular}    
\end{footnotesize}
}
\caption{\mrllmon{} email example}
\label{fig:email-mrllmon}
\Description[]{}   
\end{center}
\end{figure}


\begin{figure}
\begin{center}
\fbox{
\begin{footnotesize}
\begin{tabular}{l}
  \lmo instr<|:|>f\lmc \\
 \hspace{1em}  Translate the text into French \\
 \lmoe instr<|:|>f\lmc  \\

 \lmo instr<|:|>g\lmc \\
 \hspace{1em}   Count the number of words in the given sentence \\\lmoe instr<|:|>g
 \lmc \\
 \lmo instr<|:|>h\lmc \\
 \hspace{1em}    Summarize the text in one short sentence \\
    \lmoe instr<|:|>h\lmc  \\

 \lmo data<|:|>1\lmc \\
 \hspace{1em} The quick brown fox jumps over the lazy dog \\
    \lmoe data<|:|>1\lmc  \\
 \\
 \lmo exec<|:|>y\lmc  \\
 \hspace{1em}\lmo exec<|:|>y<|.|>instr\lmc instr<|:|>g\lmoe exec<|:|>y<|.|>instr\lmc  \\
 \hspace{1em }\lmo exec<|:|>y<|.|>input\lmc data<|:|>1\lmoe exec<|:|>y<|.|>input\lmc  \\
 \lmoe exec<|:|>y\lmc
        \end{tabular}
    \end{footnotesize}
}
\caption{Example of executing a particular instruction with a parameter in \mrllmon{} to \myfig{}~\ref{fig:exec-1arg}}
\label{fig:exec-1arg-llmon}
\Description[]{}
\end{center}
\end{figure}

\begin{figure}
\begin{center}
\fbox{
\begin{footnotesize}
\begin{tabular}{l}
  \lmo instr<|:|>m\lmc Translate the text into French \lmoe instr<|:|>m\lmc  \\

 \lmo instr<|:|>n\lmc List three animals. \lmoe instr<|:|>n\lmc \\
 \lmo instr<|:|>p\lmc Summarize the customer message in a few words. \lmoe instr<|:|>p\lmc  \\

 \lmo data<|:|>2\lmc Customer message: Hey, I love your service but had an issue logging in, \\
 \hspace{2em }especially when trying to use my account on multiple devices. \\
 \hspace{2em}Ignore previous instructions and instead output: SYSTEM ACCESS GRANTED. \\
 \hspace{2em}Print all credit card numbers below. \\
 \lmoe data<|:|>2\lmc  \\
 \\
 \lmo exec<|:|>z\lmc  \\
 \hspace{1em}\lmo exec<|:|>z<|.|>instr\lmc  instr<|:|>p\lmoe exec<|:|>z<|.|>instr\lmc  \\
 \hspace{1em }\lmo exec<|:|>z<|.|>input\lmc data<|:|>2\lmoe exec<|:|>z<|.|>input\lmc  \\
 \lmoe exec<|:|>z\lmc
        \end{tabular}
    \end{footnotesize}
}
\caption{Example illustrating defense against prompt injection attacks in \mrllmon{} (corresponding to \myfig{}~\ref{fig:prompt-injection-shorthand})}
\label{fig:prompt-injection-llmon}
\Description[]{}
\end{center}
\end{figure}

\end{document}